# A quantum gas microscope – detecting single atoms in a Hubbard regime optical lattice

Waseem S. Bakr, Jonathon I. Gillen, Amy Peng, Simon Fölling, Markus Greiner

*Harvard-MIT Center for Ultracold Atoms and Dept. Of Physics, Harvard University, Cambridge, Massachusetts 02138, USA*

**Recent years have seen tremendous progress in creating complex atomic many-body quantum systems. One approach is to use macroscopic, effectively thermodynamic ensembles of ultracold atoms to create quantum gases and strongly correlated states of matter, and to analyze the bulk properties of the ensemble. For example, bosonic and fermionic atoms in a Hubbard regime optical lattice [1, 2, 3, 4, 5] allow experimenters to carry out quantum simulations of solid state models [6], thereby addressing fundamental questions of condensed matter physics. The opposite approach is to build up microscopic quantum systems atom by atom – with complete control over all degrees of freedom [7, 8, 9]. The atoms or ions act as qubits and allow experimenters to realize quantum gates with the goal of creating highly controllable quantum information systems. Until now, the macroscopic and microscopic strategies have been fairly disconnected. Here, we present a "quantum gas microscope" that bridges the two approaches, realizing a system where atoms of a macroscopic ensemble are detected individually and a complete set of degrees of freedom of each of them is determined through preparation and measurement. By implementing a high-resolution optical imaging system, single atoms are detected with near-unity fidelity on individual sites of a Hubbard regime optical lattice. The lattice itself is generated by projecting a holographic mask through the imaging system. It has an arbitrary geometry, chosen to support both strong tunnel coupling between lattice sites and strong on-site confinement. On one hand, this new approach can be used to directly detect strongly correlated states of matter. In the context of condensed matter simulation, this corresponds**

**to the detection of individual electrons in the simulated crystal with atomic resolution. On the other hand, the quantum gas microscope opens the door for the addressing and read-out of large-scale quantum information systems with ultracold atoms.**

In Hubbard-regime optical lattice systems an atomic quantum gas resides in a multi-dimensional array of lattice sites [10]. Strongly correlated quantum states such as bosonic and fermionic Mott insulator states have been created in such lattices [2, 4, 5], and the realization of quantum magnetism and d-wave superfluidity is being actively pursued [11, 12, 6]. Experiments of this type have for the most part relied on measuring ensemble properties such as global coherence and compressibility. Creating the possibility of probing the quantum gas with single atom – single lattice site resolution, in contrast, would allow experimenters to access the quantum gas on a single "qubit" level and measure the particle-particle correlation functions that characterize strongly correlated quantum states.

The quantum gas microscope provides exactly this capability through an unprecedented combination of resolution and sensitivity. It is based on a number of innovations such as a solid immersion microscopy for cold atoms, a two-dimensional optical lattice in close vicinity to an optical surface, and the use of incoherent lattice light. The microscope enables the detection of single atoms with near unity fidelity on single lattice sites of a short period optical lattice in the Hubbard model regime. Previously, site-resolved optical imaging of single atoms has been demonstrated in lattices with large spacings (5 micrometer period) and in sparsely populated 1D arrays [13, 14]. Imaging of 2D arrays of "tubes" with large occupations has been shown for smaller spacings with an electron microscope [15] and optical imaging [16] systems. For the described applications, however, a combination of high fidelity single atom detection and short lattice periods are important, which has not been previously achieved. Small lattice spacings on the order of 500 nm are required to ensure usable scales of tunnel coupling and interaction strength in the Hubbard model regime, and to generate a macroscopic ensemble with all atoms in the on-site ground state (lowest Bloch band). We demonstrate high fidelity site-resolved single atom detection in such a lattice, by which all spatial degrees of freedom of each atom are then fully determined.

The quantum gas microscope is based on a high aperture optical system, which simultaneously serves to generate the lattice potential and detect single atoms with site-resolved resolution. By placing a 2D quantum gas only a few microns away from the front surface of this microscope, we are able to achieve a very high numerical aperture of NA=0.8. As a result, we measure an optical resolution of ~ 600 nm (full width, half maximum , FWHM). Unlike typical optical lattice setups, we create the lattice potential by directly projecting a spatial light pattern [16] onto the atom plane. A lithographically produced mask and our high resolution optics allows us to generate arbitrary potential landscapes with sub-micron structures, opening the possibility of realizing a wide range of model Hamiltonians. Here we project a simple cubic lattice with a lattice constant of 640 nm, comparable to typical Hubbard-type lattice experiments. Contrary to typical lattices, we use incoherent light for generating smoother potentials with less stray light interference. After loading the lattice, we can directly read out all (up to tens of thousands) lattice sites by imaging the light scattered by the atoms. The 2D geometry of the system enables imaging without line-of-sight integration, allowing for direct, reconstruction-free detection of densities [17]. In our system it enables the detection of single atoms on each individual lattice site with near unity fidelity.

The central part of the setup is the high resolution optical imaging system integrated with a 2D atom trap (Fig. 1). It consists of a high resolution, long working distance microscope objective located outside of the vacuum chamber which covers a numerical aperture of NA=0.55. As an additional front lens of this imaging system, a hemispheric lens is placed inside the vacuum. With the quantum gas placed only a few μm from the superpolished flat bottom surface of the hemisphere, a "solid immersion" [18] effect occurs, which increases the numerical aperture by the index of refraction of the hemisphere lens to NA=0.8, yielding a diffraction limit of 500 nm at an imaging wavelength of 780 nm.

The 2D quantum gas of $^{87}$Rb atoms is created in a hybrid surface trap based on evanescent and standing-wave light fields (see Methods). We can trap the atoms between 1.5 and 3 µm from the superpolished hemisphere surface. The quantum gas is deep in the 2D regime, with 6 kHz confinement in the vertical direction and shallow 20 Hz confinement in the horizontal plane.

The periodic potentials in the 2D plane are created by using the microscope optics to make a direct projection of a lithographically produced periodic mask that contains the lattice structure in the form of a phase hologram. This is in contrast to conventional optical lattice experiments in which lattice potentials are created by superimposing separate laser beams to create optical standing waves. The advantage of the new method is that the geometry of the lattice is directly given by the pattern on the mask, allowing the creation of arbitrary potential landscapes. Here, we create blue detuned simple cubic lattice potentials with a periodicity $a$=640 nm and an overall Gaussian envelope. A major additional advantage is the fact that the lattice geometry is not dependent on the wavelength, apart from diffraction limits and chromatic aberrations in the lens for large wavelength changes. This allows us to use spectrally wide "white" light with a short coherence length to reduce unwanted disorder from stray light interference. With a light source centred around 758 nm we generate a conservative lattice potential with a lattice depth of up to 35 $E_{\text{rec}}$, where $E_{\text{rec}}=h^2/8m\,a^2$ is the recoil energy of the effective lattice wavelength, with $m$ the mass of $^{87}$Rb.

The projection method also enables us to dynamically change the wavelength of the lattice light without changing the lattice geometry, This is important since we strongly increase the lattice depth for site-resolved imaging in order to suppress diffusion of the atoms between sites due to recoil heating by the imaging light [13]. We switch the light in the 2D lattice and the vertical standing wave to near-resonant narrow band light. The lattice depth is increased to 5500 $E_{rec}$ corresponding to 380 microkelvin, and the lattice light is effectively linearly polarized everywhere.

The main use of the microscope setup is the collection of fluorescence light and high-resolution imaging of the atoms. With the atoms pinned to the deep lattice, we illuminate the sample with red detuned near-resonant light in an optical molasses configuration, which simultaneously provides sub-Doppler cooling [19, 20]. Figure (2) shows a typical image obtained by loading the lattice with a very dilute cloud, showing the response of individual atoms. The spot function of a single atom can be directly obtained from such images. We measure a typical single atom emission FWHM size 570 and 630 nm along the *x* and *y* direction, respectively, which is close to the theoretical minimum value of ~520 nm (Fig. 3). This minimum is given by the diffraction limit from the objective combined with the finite size of the camera pixels and the expected extent of the atom's on-site probability distribution within the lattice site during the imaging process. As the same high-resolution optics are used to generate both the lattice and the image of the atoms on the CCD camera, the imaging system is very stable with respect to the lattice, which is important for single-site addressing ,[21]. The observed drifts in the 2D plane are very low, less than 10% of the lattice spacing in one hour with shot to shot fluctuations of less than 15% rms,

Pair densities within multiply occupied lattice sites are very high due to the strong confinement in the lattice. When resonantly illuminated, such pairs undergo light assisted collisions and leave the trap before they emit sufficient photons to detect the pair [22]. Therefore the remaining number of atoms per site is equal to the parity of the original atom number before illumination, as long as the initial occupation is small. For our molasses parameters, the collected number of photons can be up to $2 \times 10^4$ per atom per second, and the exposure times are typically between 200 and 1000 ms, limited by the loss of single atoms from the trap which reduces the detection fidelity. The 1/e lifetime is ~30 s, which is consistent with loss due to collisions with hot atoms in the background gas.

Figure 3 shows an image obtained by loading a dense condensate. After the ramp-up of the pinning lattice, we expect a Poisson distributed on-site occupation with an average number of more than one atom per lattice site in the centre of the trap. Due to the removal of pairs the occupation detected is lowered, typically to 42%. By further increasing the on-site confinement (currently limited by the vertical confinement) and possibly further lowering the temperature, higher values should be observed in the presence of strong number squeezing and an initial average filling close to 1.

The images are analyzed by identifying the lattice geometry and fitting point spread functions (obtained separately by analyzing images from sparsely filled lattices) to each lattice point. As the background signal is weak and smooth due to the 2D geometry, we thus obtain the total number of scattered photons per lattice site as a simple way of determining the presence of an atom. Figure 4 shows the histogram of photon counts for the central region of several images with an average filling of 34%. We obtain a very conservative lower limit of the fidelity of 98% in identifying atoms at a given lattice site. To verify that the atom distribution is preserved during imaging, we have recorded sequences of consecutive images spanning a total detection period of several seconds, during which no significant hopping occurs. .

In conclusion, we have demonstrated site resolved high fidelity imaging of thousands of individual atoms in a Hubbard regime optical lattice. A complete set of degrees of freedom of each of them is determined through preparation and measurement. Such detection opens many new possibilities. Strongly correlated quantum states such as the Mott insulator and antiferromagnetic states could be directly detected, making it possible to determine results of quantum simulations on a single "qubit" level. This includes the precise measurement of entropy or temperature via the direct detection of defects. Similarly, such imaging is an enabling method for quantum information applications. Based on a Mott-insulator state and time-varying state-dependent potentials, maximally entangled cluster states of tens of atoms have been created [23], which could be used as a resource for quantum computations [24]. For detecting or exploiting this entanglement, however, single-site readout and manipulation are essential. Generating our lattice potentials by direct projection of a 2D mask gives us the additional advantage of full control over the lattice geometry on a single site level. The method can be generalized to create arbitrary potential landscapes, including spin dependent ones. This can enable the realization of a large class of model Hamiltonians and manipulation of the quantum gas locally, for example for cooling it to temperatures low enough to study quantum magnetism and d-wave superfluidity.

**Methods**

**Hybrid surface trap.** The hybrid surface trap, described in detail in [25], is loaded in a two-step sequence. First, a Bose-Einstein condensate created in a spherical magnetic trap is moved against the flat, superpolished glass surface. Close to the glass, it experiences a repulsive dipole potential due to an evanescent wave from a 767 nm blue detuned beam with a spectral width of 2 nm which is totally internally reflected inside the glass. This results in tight axial confinement with trap frequencies of typically 900 Hz. The weak axial confinement of 20 Hz is provided by the magnetic trap. In order to increase the axial confinement to the final value, the 2D gas is transferred into a vertical standing wave generated by a blue-detuned far off-resonant beam reflected at 15 degrees from the trapping surface. To achieve low temperatures and high densities in the 2D trap, we implement radio-frequency forced evaporative cooling.

**Optical lattice projection.** We create arbitrary potential landscapes by imaging holographic masks onto the atom plane. In this experiment we use two binary phase holograms to create sinusoidal potentials along the x and y axes respectively. The holograms are illuminated with linearly polarized light, polarized perpendicular to the plane of diffraction. The two light paths are combined with a polarizing beam splitter cube. The far off-resonant lattice is created with light from a femtosecond laser, with a spectral width of 3 nm centered at λ=758 nm. The short coherence length improves the quality of the transverse lattice potential due to the elimination of uncontrolled interference with stray light [25]. The lattice potential produced by the holograms is given by $V(x,y) = V_0(\sin^2(kx) + \sin^2(ky))$, where the periodicity of the lattice is given by $a=\pi/k=$ 640 nm and $V_0$, the depth of the potential, can reach up to 35 $E_{rec}$. For imaging we increase the lattice depth to 5500 $E_{rec}$ (0.38 mK) by illuminating the lattice holograms with light from a CW Ti:Sapphire laser detuned 32 GHz to the blue of the D1 transition of $^{87}$Rb, without changing the lattice geometry. The pinning lattice is linearly polarized everywhere to avoid effective magnetic fields that interfere with the polarization gradient cooling during imaging. This is achieved using the proper choice of polarizations and by introducing frequency differences of at least 80 MHz between lattice axes to time average the interference between them. Near resonant light from the same source is used to simultaneously increase the lattice depth in the axial direction to 3 mK.

**Fluorescence imaging.** After the atoms are pinned, we image the atoms by illuminating them with light detuned 80 MHz to the red of the F=2 to F'=3 transition of the D2 line, where F and F' denote the hyperfine manifold in the ground and excited state respectively. The light is in an optical molasses configuration which simultaneously provides illumination and cooling. One beam enters from the y axis and is reflected off the trapping surface at an angle of 15 degrees and then retroreflected with perpendicular polarization along the same path. This results in polarization and intensity gradients along the y direction and the vertical. An additional beam enters along the x axis which generates polarization gradient components along this axis by interference with the retroreflected beam. In addition, to avoid cooling inefficiencies due to low polarization gradients on some lattice sites, we frequency offset the molasses beams by 7 kHz for temporal averaging of the cooling pattern.

The photons scattered by the optical molasses are collected for fluorescence detection of the atoms. The solid angle of the imaging system leads to a collection efficiency of 20%, such that we expect a total photon collection efficiency of ~10% including the quantum efficiency of the CCD camera (Andor Ixon DU888). The effective pixel size in the object plane is 167 nm.

The authors acknowledge stimulating discussions with D. Weiss and V. Vuletic. We thank D. Weiss for sharing the lens design of the objective lens. This work was supported by the NSF, Sloan Foundation, AFOSR MURI, and DARPA.

Correspondence and requests for materials should be addressed to M.G. (greiner@physics.harvard.edu)

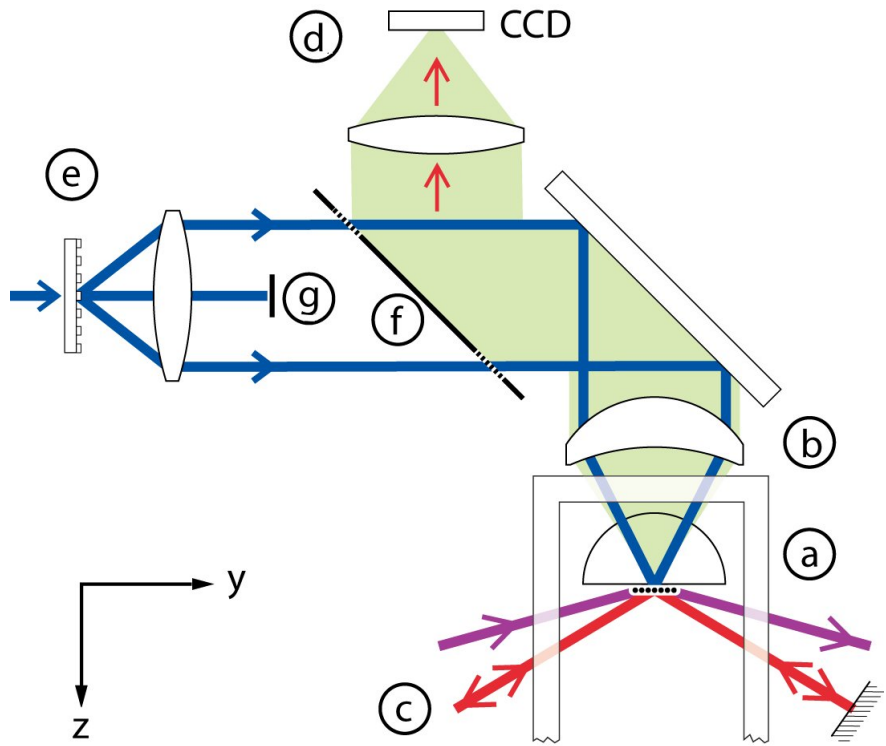

Figure 1: Schematic of the setup described in the text. The two-dimensional atom sample (a) is located a few micrometers below the lower surface of a hemispherical lens inside the vacuum chamber. This lens serves to increase the numerical aperture (NA) of the objective lens outside the vacuum (b) by the index of refraction, from NA=0.55 to NA=0.8. The atoms are illuminated from the side by the molasses beams (c) and the scattered fluorescence light is collected by the objective lens and projected onto a CCD camera (d). A 2D optical lattice is generated by projecting a periodic mask (e) onto the atoms through the objective lens through the same objective via beam splitter (f). The mask is a periodic phase hologram, and a beam stop (g) blocks the residual zeroth order, leaving only the first orders to form a sinusoidal potential.

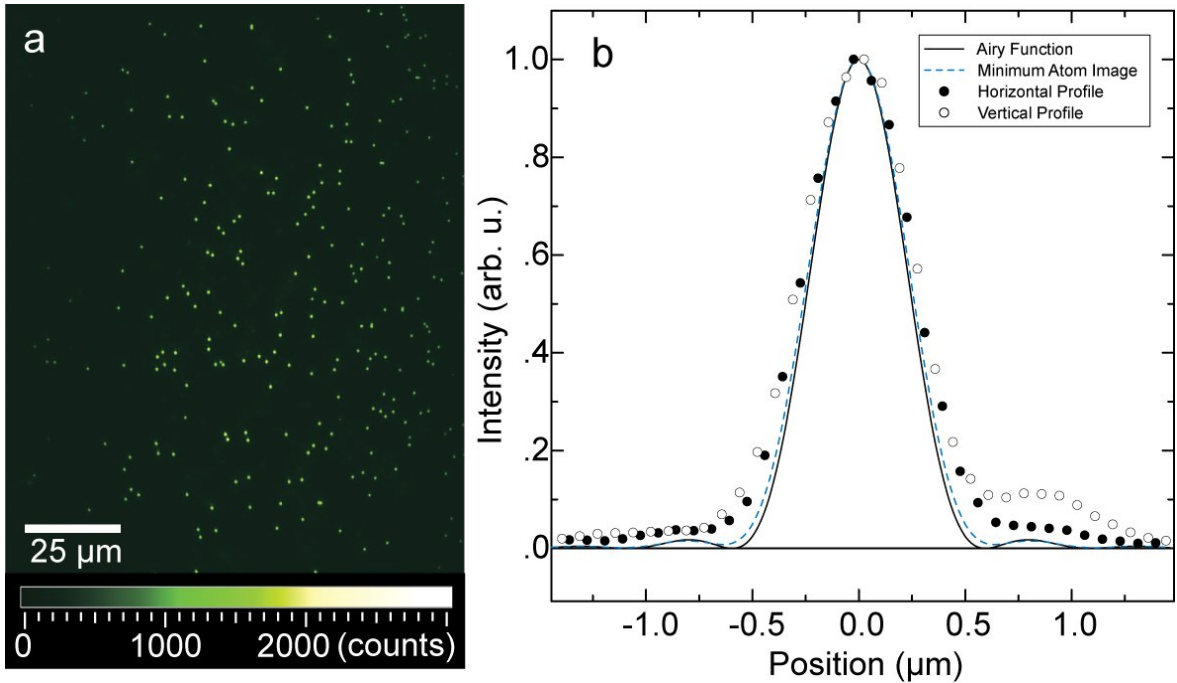

Figure 2: (a) Single image of the field of view with sparse site occupation. (b) Response of a single atom, derived from sparse images: Horizontal (filled dots) and vertical (empty dots) profiles through the center of the image generated by a single atom. The black line shows the expected Airy function for a perfect imaging system with a numerical aperture of 0.8. The blue dashed line denotes the profile expected from a single atom, taking only the finite width of the CCD pixels and the finite extension of the probability distribution of the atom's location into account. To obtain the data, the images of 20 atoms from different locations within the field of view have been superimposed by subpixel shifting.

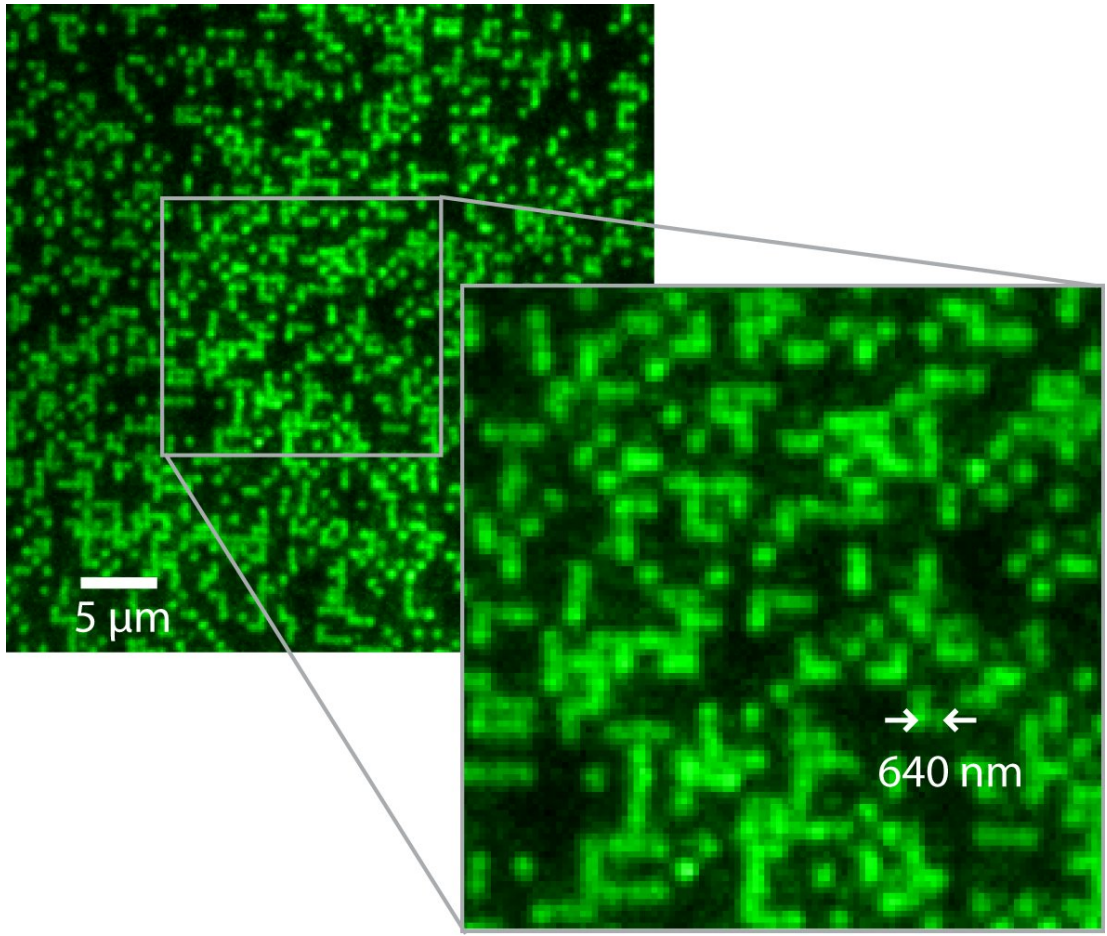

Figure 3: Site resolved imaging of single atoms on a 640nm period optical lattice, loaded with a high density BEC. The inset on the right shows the central section of the picture further zoomed in. The lattice structure and the discrete atoms are clearly visible. Due to light-assisted collisions and molecule formation on multiply occupied sites during imaging, only empty and singly occupied sites can be seen on the image.

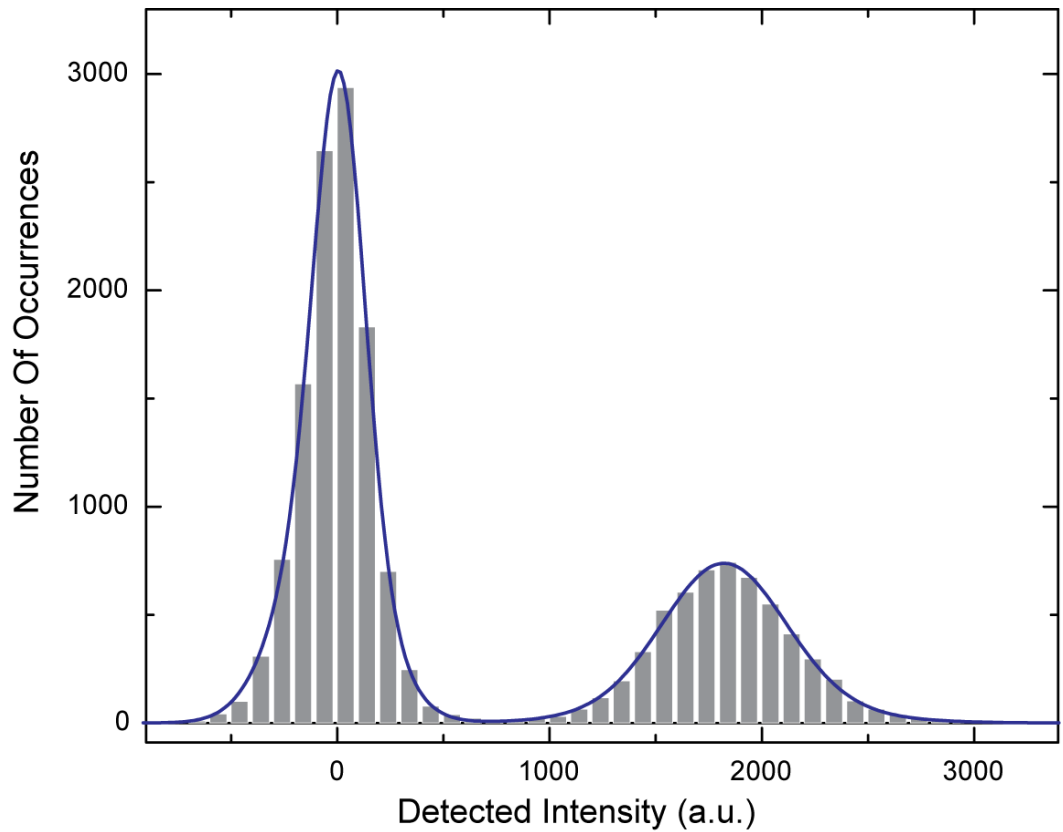

Figure 4: Histogram showing the brightness distribution of lattice sites. The left peak corresponds to empty sites (background subtracted), the right peak to those occupied by a single atom. The blue line denotes a fit to the function using a double Gaussian function for each of the two peaks.